\documentclass[conference]{IEEEtran}
\IEEEoverridecommandlockouts

\makeatletter
\def\ps@headings{%
\def\@oddhead{\mbox{}\scriptsize\rightmark \hfil \thepage}%
\def\@evenhead{\scriptsize\thepage \hfil \leftmark\mbox{}}%
\def\@oddfoot{}%
\def\@evenfoot{}}
\makeatother
\usepackage{cite}
\usepackage{amsmath,amssymb,amsfonts}
\usepackage{algorithmic}
\usepackage{graphicx}
\usepackage{tikz}
\usetikzlibrary{positioning}
\usepackage{gensymb}
\usepackage{caption}
\usepackage{subcaption} 
\usepackage{textcomp}
\usepackage{xcolor}
\def\BibTeX{{\rm B\kern-.05em{\sc i\kern-.025em b}\kern-.08em
    T\kern-.1667em\lower.7ex\hbox{E}\kern-.125emX}}

\usepackage[T1]{fontenc}
\usepackage[utf8]{inputenc}
\usepackage{authblk}

\usepackage{tikz}
\newcommand\copyrighttext{%
  \footnotesize © 2024 IEEE. Personal use of this material is permitted. Permission from IEEE must be obtained for all other uses, in any current or future media, including \\ reprinting/republishing this material for advertising or promotional purposes, creating new collective works, for resale or redistribution to servers or lists, or reuse of any copyrighted component of this work in other works.}

\newcommand\copyrightnotice{%
\begin{tikzpicture}[remember picture,overlay]
\node[anchor=south,yshift=10pt] at (current page.south) {\fbox{\parbox{\dimexpr\textwidth-\fboxsep-\fboxrule\relax}{\copyrighttext}}};
\end{tikzpicture}%
}

\begin{document}
\title{IoTWarden: A Deep Reinforcement Learning Based Real-time Defense System to Mitigate Trigger-action IoT Attacks}


\author[1]{Md Morshed Alam}
\author[2]{Israt Jahan}
\author[1]{Weichao Wang}
\affil[1]{Department of Software and Information Systems, University of North Carolina at Charlotte, Charlotte, USA}
\affil[2]{Department of Computer Science, University of Memphis, Memphis, USA} 
\affil[ ]{\{malam3, wwang22\}@uncc.edu, ijahan1@memphis.edu}


\maketitle
\copyrightnotice

\begin{abstract}
In trigger-action IoT platforms, IoT devices report event conditions to IoT hubs notifying their cyber states and let the hubs invoke actions in other IoT devices based on functional dependencies defined as rules in a rule engine. These functional dependencies create a chain of interactions that help automate network tasks. Adversaries exploit this chain to report fake event conditions to IoT hubs and perform remote injection attacks upon a smart environment to indirectly control targeted IoT devices. Existing defense efforts usually depend on static analysis over IoT apps to develop rule-based anomaly detection mechanisms. We also see ML-based defense mechanisms in the literature that harness physical event fingerprints to determine anomalies in an IoT network. However, these methods often demonstrate long response time and lack of adaptability when facing complicated attacks. In this paper, we propose to build a deep reinforcement learning based real-time defense system for injection attacks. We define the reward functions for defenders and implement a deep Q-network based approach to identify the optimal defense policy. Our experiments show that the proposed mechanism can effectively and accurately identify and defend against injection attacks with reasonable computation overhead.     

\end{abstract}


\section{Introduction}
Emerging Internet of Things (IoT) platforms support \textit{trigger-action} functionality that enables IoT hubs in smart homes instructing IoT devices to perform predefined actions (e.g., \textit{turning on smart light}) based on specific event conditions reported by other IoT devices (e.g., \textit{the door is unlocked}) \cite{ozmen2023-evation-attacks}. Hence, event conditions act as \textit{triggers}, which activate corresponding actions associated with rules set in a \textit{rule engine} defined by users in the IoT hubs. These rules represent functional dependencies between various IoT event conditions and actions. In a smart home network, these functional dependencies create a chain of interactions among IoT devices to automate network tasks. IoT devices report event conditions to the hub informing the cyber states of the devices themselves, and the hub leverages physical evidence captured by some deployed sensors in the network to verify the physical states of those devices. The trigger-action functionality also allows the hub to communicate with users with important notifications and alerts about the physical developments (e.g., \textit{when the fire alarm sounds}) in a smart home environment \cite{celik2019a_iotsec_survey}. SmartThings \cite{smartthings2022} and IFTTT \cite{IFTTT2022} are two prominent examples of such platforms. 


Although the chain of interactions generated in a smart home enables network automation, it creates security vulnerabilities in the environment \cite{alam2022_iotmonitor}. Attackers can exploit this chain to manipulate IoT devices to trigger sensitive actions in the network, such as setting up thermostat at 120\degree F. Attackers can also collect seamless sensitive user data without raising suspicion to the defense system, or cause unsafe state transitions in the environment violating rule execution integrity \cite{fan2021_ruleedger}. In this paper, we focus on the \textit{remote injection} attack \cite{alam2022_iotmonitor} that allows attackers to inject fake event conditions into the hub and trick it to command target IoT devices to perform some actions that help attackers achieve certain attack goals. For instance, given that there is a functional dependency between fire alarm and smart lock, an attacker can inject a fake fire alarm event condition to the IoT hub representing a false fire hazard situation and let the hub instruct the lock to automatically open the front door. This attack is also called \textit{event spoofing} attack since attackers deceive the hub with malicious reporting of spoofed event conditions \cite{ozmen2023-evation-attacks}. Thus, it is possible for attackers to exploit functional dependencies present in the chain of interactions to implement safety-critical attacks in smart environments \cite{alam2021_survey}.


In the literature, we see great efforts to address this security vulnerability. For example, we see rule-based and ML-based IoT anomaly detection systems (also called \textit{event verification} systems) that use \textit{physical event fingerprints} of IoT devices to verify whether the reported events to the hub occurred physically \cite{sbirnbach2019_peeves} \cite{fu2021-hawatcher}. Sensors deployed in the environment to continuously measure physical channels of devices (e.g., \textit{light intensity}) help determine these unique fingerprints. We also see approaches that diagnose unsafe and insecure state transitions between IoT devices in the network to ensure that each event reporting to the hub complies with the defined security policy of the network \cite{leonardo2018_iotdots} \cite{celik2019b_iotguard}. However, most of the existing security approaches either require active intervention on the IoT app source code, or do not provide realtime security against an ongoing trigger-action attack.

 
In this paper, we propose \textit{IoTWarden}, a deep reinforcement learning (deep RL) based defense system that allows a defense agent to model attack behavior based on the impact of attack actions in the IoT environment and obstructs the progression of an ongoing attack in realtime. We make the following contributions in this paper:

\begin{itemize}
    \item We propose a deep RL based realtime defense system, namely \textit{IoTWarden}, that allows defenders to take necessary defense actions at runtime against ongoing trigger-action attacks.

    \item We implement an LSTM-based \cite{LSTM_original_paper_Hochreiter1997} Recurrent Neural Network (RNN) to discern optimal attack sequences to help IoTWarden infer attack behavior at runtime.

    \item We implement Deep Q-Network \cite{drl_mnih2013playing} to obtain optimal defense policy and train the decision process of defenders. 

    \item We conduct extensive experiments and simulation to evaluate the performance of IoTWarden, showing that the adoption of optimal defense policy yields improved security gain with very low computation overhead.



\end{itemize}

The rest of the paper is organized into six sections. In section II, we introduce the trigger-action attack, discuss attack strategy, and define a relevant threat model. We also describe how defenders characterize the attack progression by generating attack graphs. In section III, we present the IoTWarden defense system and describe each individual component in detail. We provide the details of our experiment and simulation in section IV. Later, in section V, we include the simulation results and evaluate the performance of the system in countering attacks in realtime. Finally, we conclude the paper summarizing the methodology and discuss future extensions in section VI. 

\section{Attack Overview}

To perform a remote injection attack exploiting functional dependencies in a smart home network, an attacker injects necessary amount of event conditions into the hub, which act as \textit{triggers} for the actions in other devices. We call the injection operations \textit{exploits}, which trick the hub to activate actions in target devices. Note that the terms \textit{event condition} and \textit{event} signify the same aspect in our attack scenario, and therefore, we interchangeably use these terms throughout the rest of the paper to avoid verbosity.


\subsection{Attack Definition}
We assume that there are $\mathcal{N}$ possible event conditions in the network $C \negmedspace = \negmedspace \{c_i\}, 1 \negmedspace \leq \negmedspace i \negmedspace \leq \negmedspace \mathcal{N}$, and the attacker conducts an optimal sequence of $\mathcal{M}$ exploits $\xi \negmedspace = \negmedspace \{e_j\}, 1 \negmedspace \leq \negmedspace j \negmedspace \leq \negmedspace \mathcal{M}$ to report a subset of these conditions to the hub to perform a trigger-action attack. We assume that the attacker performs an exploit $e_t \negmedspace \in \negmedspace \xi$ at timestep $t$ to maliciously report an event condition $c_t \negmedspace \in \negmedspace C$ to the hub and expects the hub to invoke an event condition $c_{t'} \negmedspace \in \negmedspace C$ at another device, which is a target of the attacker. Since complex functional dependencies may exist in the chain of interactions, it is possible that the attacker needs to perform a set of exploits $\{e_t\} \negmedspace \in \negmedspace \xi$ at timestep $t$ to report multiple event conditions $\{c_t\} \negmedspace \in \negmedspace C$ to achieve the attack goal. The attacker considers an IoT device compromised if the following definition 1 applies to it.


\textbf{\textit{Definition 1 (IoT Device Compromise):}}
An IoT device $d_g$ is said to be \textit{compromised} at timestep $t$ if an event condition $c_g \negmedspace \in \negmedspace C$ is triggered at the device $d_g$ due to the malicious reporting of a set of event conditions $\{c_{t}\} \negmedspace \in \negmedspace C$ to the IoT hub by an attacker through a sequence of \textit{exploits} $\{e_{t}\} \negmedspace \in \negmedspace \xi $ adopting a strategy $z \negmedspace \in \negmedspace Z$ that dictates the attack operations.

\subsection{Attack Characterization}
To model complex interactions of the attacker with the network and characterize the attack progression over time, we assume that the defense agent embeds event conditions and corresponding exploits to construct a \textit{directed acyclic attack dependency graph} $G \negmedspace = \negmedspace \{C, \xi\}$. Since attack dependency graph grows non-linearly across time horizon with the increase of the number of nodes, the defense agent considers the inclusion of \textit{monotonicity} \cite{ammann2022-monotonicity} property in attack behavior to prevent \textit{state explosion problem}  \cite{sheyner2002-attack-graph-scalability-issue} and keep the attack dependency graph reasonably small to perform security analysis. This property enforces a constraint on the attack behavior by limiting the influence of past exploits on the future ones. We assume that the defender constructs this graph using state-of-the-art vulnerability analysis tools, such as TVA tool of \cite{jajodia2006-attack-graph-generation}.

\subsection{Attack Strategy}
Since exploits drive the remote injection attack, it is important for the attacker to follow an effective attack strategy in choosing the optimal exploits. We assume that the attacker crafts a set of attack strategies $Z \negmedspace = \negmedspace \{z_i\}, 1 \negmedspace \leq \negmedspace i \negmedspace \leq \negmedspace \kappa$, where $\kappa$ is the total number of possible strategies for the attacker. Hence, each $z \negmedspace \in \negmedspace Z$ gives a unique set of exploits to perform during the attack. The attacker changes the adopted attack strategy once the defender blocks certain event conditions making some event reporting untenable. We assume that the attacker precomputes optimal attack paths from each IoT device to the ultimate target device using LSTM-based RNNs exploiting temporal dependencies exist in the network. Based on these optimal attack paths, the attacker crafts several exploit sequences, which act as attack strategy $z \negmedspace \in \negmedspace Z$.

\subsection{Threat Model}
We assume that the attacker injects event conditions into the IoT hub using multiple ghost devices with stolen IDs and credentials of valid devices. A ghost device usually is a computer program designed to simulate the behavior of a real IoT device. The attacker steals information about valid devices from public data hosted in manufacturer's websites or public repository, including GitHub. The attacker also extracts real-time event information from network traffic using different IoT network analysis tools \cite{Trimananda2020PacketLevelSF} \cite{zhang2018_homonit}. We also assume that the attacker is intelligent enough to profile the impact of IoT event occurrences over the physical environment and perform opportunistic attack similar to the attacks explained in \cite{sbirnbach2019_peeves}. To evade detection by the defense system, the attacker follows a minimally invasive attack strategy that requires as few injections as possible. From a limited pool of attack strategies, the attacker dynamically selects strategy by observing defender's actions in realtime. As we discuss in section I, the defense system uses physical evidence about the network collected by sensors to verify event occurrences in the network. We assume that the attacker cannot compromise these sensors and is incapable of compromising the IoT hub through which all sorts of communication among IoT devices occur in a smart home network. 

\section{IoTWarden Defense System}
We design \textit{IoTWarden} as a real-time defense system that infers attack behavior by monitoring the impact of attacker's actions in the network and adopts an optimal policy to select defense actions, maximizing the total security reward. IoTWarden continuously assesses the security status of the smart home network and takes necessary defense actions to make some exploits infeasible so that the attacker cannot complete a trigger-action IoT attack. We assume that IoTWarden is hosted in the hub, and it enables the hub to block the activation of actions in target IoT devices when it identifies the event injection in the network. 

\begin{figure}[!h]
    \centering
    \captionsetup{justification=centering}
    \includegraphics[scale=0.40]{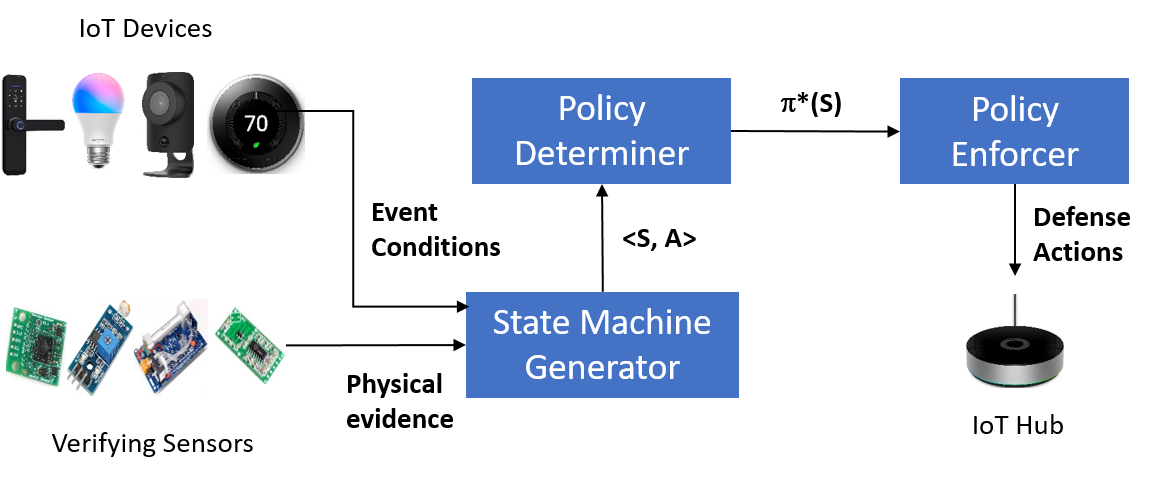}
    \caption{IoTWarden System Architecture}
    \label{fig:iotwarden-sysyem-architecture}
    \vspace{-10pt}
\end{figure}

IoTWarden consists of three main components: 1) state machine generator, 2) policy determiner, and 3) policy enforcer. Fig. \ref{fig:iotwarden-sysyem-architecture} presents the system architecture of IoTWarden.

\subsection{State Machine Generator}

Considering the IoT network as system environment, this component creates a \textit{finite state machine} with $N$ unique system states, such as $S \negmedspace = \negmedspace \{s_i\}, 1  \negmedspace \leq  \negmedspace i  \negmedspace \leq  \negmedspace N$ and $M$ unique actions, such as $A \negmedspace = \negmedspace \{a_k\}, 1  \negmedspace \leq  \negmedspace k  \negmedspace \leq  \negmedspace M$. Hence, $s_t \negmedspace \in \negmedspace S$ represents the environment state inferred by the hub at timestep $t$ due to the reporting of $\{c_t\} \negmedspace \in \negmedspace C$, and the action $a_t \negmedspace \in \negmedspace A$ makes the environment transition into another state $s_{t+1}$ at timestep $t+1$ with a probability $T(s_t, a_t, s_{t+1}) \negmedspace = \negmedspace Pr(s_{t+1} | s_t, a_t)$ yielding a security gain (reward) $R(s_t, a_t, s_{t+1}) = \{r_t\}$. We assume that the action space $A$ contains the following four $(M \negmedspace = \negmedspace 4)$ actions: \{\textit{event injection ($a_1$)}, \textit{checking device accessibility ($a_2$)}, \textit{monitoring security status of the network ($a_3$)}, and \textit{blocking triggers ($a_4$)}\}. IoTWarden chooses $<\negmedspace a_1, a_2 \negmedspace>$ to mimic the behavior of an attacker during the training phase of the system, and the defense agent takes the actions $<\negmedspace a_3, a_4 \negmedspace>$ during both the training and the deployment phase. 

\subsubsection{Reward Function}
We design the defense agent to be reactive against attack actions. If \textit{event injection} actions are taken aggressively, we want the defense agent to take \textit{blocking triggers} actions more frequently. We define a parameter \textit{attack proximity factor} $p$ to indicate how close the attack is to the ultimate goal node. We compute $p$ by taking ratio between the number of events already compromised and the total events in the attack chain. When $p$ increases, we want the defense agent to take more \textit{blocking triggers} actions. The ultimate goal here is not to let the attacker compromise the goal node. Therefore, we design the reward function to help the defense agent decide when to allow the attacker to perform injection operations and when to start blocking triggers. The reward function is defined using the following equation \eqref{eqn:reward_function}:   

\vspace{-10pt}
\begin{equation}
    R(.) =  
    \begin{cases}
        n_{a_3} r_{a_3} - \frac{p n_{a_1} r_{a_1}}{n_{a_1} + n_{a_2}}  - G_r & \mbox{ if } \frac{n_{a_1} p}{n_{a_1} + n_{a_2}} < k \\ \\
       
        \frac{(n_{a_4} r_{a_4}) (n_{a_3} r_{a_3})}{n_{a_4} + n_{a_3}} \\ - max \big(n_{a_2} r_{a_2}, \frac{p n_{a_1} r_{a_1}}{n_{a_1} + n_{a_2}} + G_r \big) & \mbox{otherwise}
    \end{cases}
    \label{eqn:reward_function}
\end{equation}

Hence, $n_{a_1}, n_{a_2}, n_{a_3}, \text{ and } n_{a_4} \negmedspace$ respectively represent the number of actions $a_1, a_2, a_3, \text{ and } a_4$ taken in the environment. On the contrary, $r_{a_1}, r_{a_2}, r_{a_3}, \text{ and } r_{a_4} \negmedspace$ respectively represent the immediate reward given by the environment for taking the actions $a_1, a_2, a_3, \text{ and } a_4$. The injection threshold $k$ is a user defined parameter here. The value of $k$ indicates the tolerance of the defense agent against the injection operations. The parameter $G_r$ represents the reward for the attacker to compromise the ultimate goal node, and the defense agent always tries to make that impossible. Any node compromised beyond goal node yields $G_r = 0$. 

\subsection{Policy Determiner}
IoTWarden adopts an optimal policy that dictates how the defense agent chooses actions at runtime. To determine this policy, IoTWarden solves a Markov Decision Process $< \negmedspace S, A, T, R \negmedspace>$ and trains a policy that maximizes the future discounted reward, $R_t \negmedspace = \negmedspace \sum_{t=0}^{t=T} \gamma^t r_t$ at timestep $t$, where the simulation stops at $t=T$, and $\gamma$ ($0 \leq \gamma \leq 1$) is the discount factor. Given a state $s \in S$, the objective here is to discern a policy $\pi(s)$ that provides optimal state-action value pairs $< \negmedspace s,a \negmedspace>$ from a function $Q^*(s, a)$ (called \textit{Q-function}) yielding the maximum security reward $R_t$, such as:

\vspace{-10pt}
\begin{equation}
    Q^{\pi^*}(s_t,a_t) = \max_\pi E \big[R_t | s_t = s, a_t = a, \pi \big]
    \label{eqn:q-function}
\end{equation}

The Q-function obeys the Bellman equation \cite{MDP_bellman1957}, and the equation \eqref{eqn:q-function} can be rewritten as:

\vspace{-15pt}
\begin{equation}
    Q^{\pi^*}(s_t,a_t) = E \big[r_t + \gamma \max_{a' \in \pi(s_t)} Q^*(s_{t+1}, a' | s_t, a) \big]
    \label{eqn-bellman-equation}
\end{equation}

The policy determiner constructs a Deep Q-Network (DQN) \cite{drl_mnih2013playing} to estimate the Q-function and utilizes a neural network function approximator with weights $\theta_i$ at the iteration $i$ to compute the temporal difference error $\delta$. As shown in equation \eqref{eqn-loss-function}, it uses the \textit{Huber loss} \cite{huber_loss_1964} function to minimize this error $\delta$ over a batch $\beta$ of transitions $<\negmedspace s_t, a_t, s_{t+1}, r_t \negmedspace>$ from a replay memory. These transitions represent the defense agent's history of interactions with the environment. The policy determiner samples these transitions from the replay memory using an \textit{exploration-exploitation trade-off} approach similar to the $\epsilon$-greedy approach discussed in \cite{epsilon-greedy-wunder-2010}. The following equation \eqref{eqn-temporal-difference} presents the formula used to compute $\delta$ and equation \eqref{eqn-loss-function} shows how this quantity $\delta$ is minimized during the training period.    

\vspace{-15pt}
\begin{equation}
    \delta = E \big[r_t + \gamma \max_{a' \in \pi(s_t)} Q(s_{t+1}, a'; \theta_{i+1}) - Q(s_t, a; \theta_{i}) \big]
    \label{eqn-temporal-difference}
\end{equation}

\begin{equation}
    \mathcal{L} = \frac{1}{|\beta|} \sum_{{(s_t, a, s_{t+1}, r_t)} \in \beta} \mathcal{L(\delta)}
    \label{eqn-loss-function}
\end{equation}
where,
\begin{equation*}
    \mathcal{L(\delta)} =   
    \begin{cases}
        \frac{1}{2} \delta^2 &\mbox{ if } |\delta| \leq 1 \\
        |\delta| - \frac{1}{2} & \mbox{otherwise}
    \end{cases}
\end{equation*}

When we train the DQN, we ensure that the following action selection method is adopted to balance between exploration and exploitation.

\begin{itemize}
    \item Exploration: an action $a$ is randomly selected from $A$ with the probability of $\epsilon$,
    \item Exploitation: the optimal action $a_t \in A$ is chosen as a greedy option with the probability of $1 - \epsilon$.
\end{itemize}

We formulate $\epsilon$ using the following equation \eqref{eqn-epsilon-determination}:
\begin{equation}
    \epsilon = \epsilon_{end} + (\epsilon_{start} - \epsilon_{end}) \medspace e^{{\frac{-t}{\epsilon_{decay}}}}
    \label{eqn-epsilon-determination}
\end{equation}

Hence, $\epsilon_{start}$ and $\epsilon_{end}$ respectively denotes the maximum and minimum value of a range in which the value of $\epsilon$ lies. The parameter $\epsilon_{decay}$ denotes the decay factor that gradually diminishes the value of $\epsilon$, and $t$ denotes the current timestep.

\subsection{Policy Enforcer}
\begin{figure}[!h]
    \centering
    \captionsetup{justification=centering}
    \includegraphics[scale=0.36]{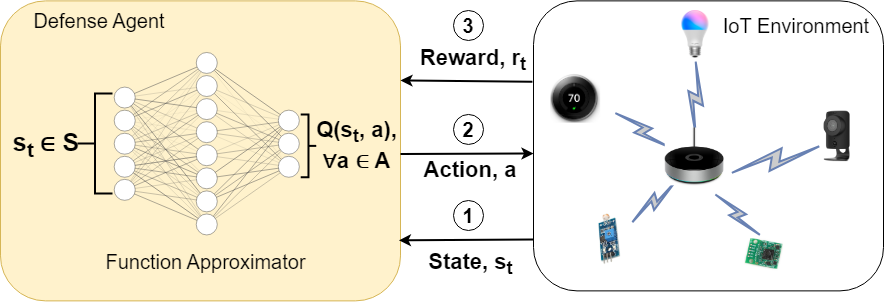}
    \caption{Illustration of IoTWarden Policy Enforcement}
    \label{fig:iotwarden-policy-enforcement}
\end{figure}

To counter a realtime attack, IoTWarden leverages the trained policy $\pi^*$ to choose optimal defense actions. At timestep $t$, IoTWarden receives the state $s_t$ from the environment and utilizes the function approximator to choose the best defense action $a_t \in A$ that maximizes the ultimate reward. Once the policy enforcer takes the chosen action, the environment returns a reward $r_t$ to the defense agent based on the reward function given in the equation \eqref{eqn:reward_function}. The complete policy enforcement process is illustrated in Fig. \ref{fig:iotwarden-policy-enforcement}. 

\section{Experiment and Simulation}
We implement IoTWarden using TensorFlow and conduct experiments simulating a trigger-action attack in a smart home network to evaluate the performance of the system. We utilize the PEEVES \cite{sbirnbach2019_peeves} dataset to extract the state space needed to be encoded in the simulating environment. This dataset contains event traces collected from 12 different IoT devices and physical evidence measured by 48 different sensors. We utilize 24-hours data in our experiment. We run the experiment on an Apple M1 Pro machine with 16 GB RAM and 8-core GPU. 

\subsection{Determining Optimal Attack Sequences}
We use a Hidden Markov Model based approach described in \cite{alam2022_iotmonitor} to determine IoT events likely to occur in a trigger-action attack. We call them \textit{crucial} nodes. Later, we implement an LSTM \cite{LSTM_original_paper_Hochreiter1997} based RNN to create optimal attack sequences exploiting temporal dependencies exist among IoT events captured in the PEEVES \cite{sbirnbach2019_peeves} dataset and considering each crucial node as the starting node of a possible attack chain. The RNN architecture contains one embedding layer to convert input event sequences into fixed-length vectors of size 128, two bidirectional LSTM layers with 64 and 32 units respectively, and one fully-connected output layer with 23 classes. Each event sequence given as input to the embedding layer records 227 different event occurrences.      




\subsection{IoT Environment}
From the optimal attack sequences, we extract events based on origin, convert them into states, and encode them into the system environment for simulation. We encode 12 unique states and 4 actions in the environment using OpenAI Gym \cite{gym_2016}. We also integrate the reward function in the environment so that the defense agent is rewarded accordingly for the actions taken at runtime.



\subsection{Function Approximator}
Given a state in the environment, we use a neural network \textit{function approximator} to estimate reward for each available action and output the optimal action yielding the highest reward. The neural network we use has 2 hidden layers with 64 and 32 neurons, while the input and output layers include 128 and 4 units respectively. We use the $ReLU$ activation function for the input and hidden layers. For the output layer, we use the $Linear$ activation function. We use \textit{Huber loss} \cite{huber_loss_1964} as the loss function in this neural network. 



\subsection{Deep Q-Network} 
The DQN we use to estimate Q-function has the hyperparameter settings listed in Table \ref{table:hyperparamter-setting}. We train the network for $250$ episodes and use \textit{Adam} \cite{adam-optimizer-2017} optimizer with the learning rate, $\alpha \negmedspace = \negmedspace 1e^{-4}$. During our training, we update a policy network at a constant rate, $\tau=20$. To record the defense agent's interaction with the environment, we use a replay buffer of size $50,000$ and sample $16$ interactions at a time when the optimal action is chosen.

\begin{table}[!h]
    \centering
    \begin{tabular}{| c | c |} 
         \hline
         \textbf{Parameter} & \textbf{Quantity} \\ \hline
         Total episodes & 250 \\ \hline
         Number of epochs per episode & 100 \\ \hline
         Optimizer & Adam \\ \hline
         Minibatch size, $\beta$ & 16 \\ \hline
         Discount factor, $\gamma$ & 0.95 \\ \hline 
         Learning rate, $\alpha$ & $1e^{-4}$ \\ \hline
         $(\epsilon_{start}, \epsilon_{end}, \epsilon_{decay})$ & (1.0, 0.1, 0.99999) \\ \hline
         Target network update frequency, $\tau$ & 20 episodes \\ \hline
         Replay buffer size & 50,000 \\ \hline
    \end{tabular}
    \caption{Hyperparameter settings for Deep Q-network}
    \label{table:hyperparamter-setting}
\end{table}

\section{Results and Evaluation}

\subsection{Optimal Attack Sequence}
\begin{figure}[!h]
    \centering
    \captionsetup{justification=centering}
    \begin{subfigure}[t]{0.24\textwidth}
        \includegraphics[width=\textwidth]{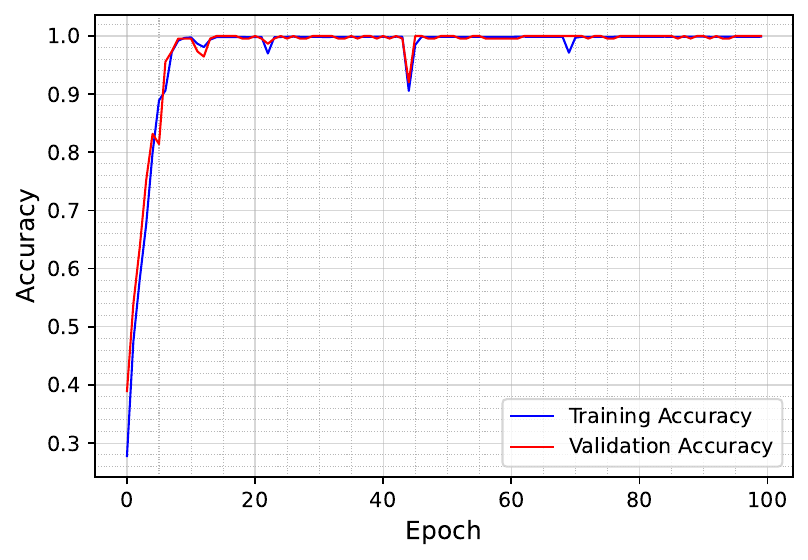}
        \caption{Training and validation accuracy over epochs}
        \label{fig:seq_gen_acc_eval}
    \end{subfigure}
    \hfill
    \begin{subfigure}[t]{0.24\textwidth}
        \includegraphics[width=\textwidth]{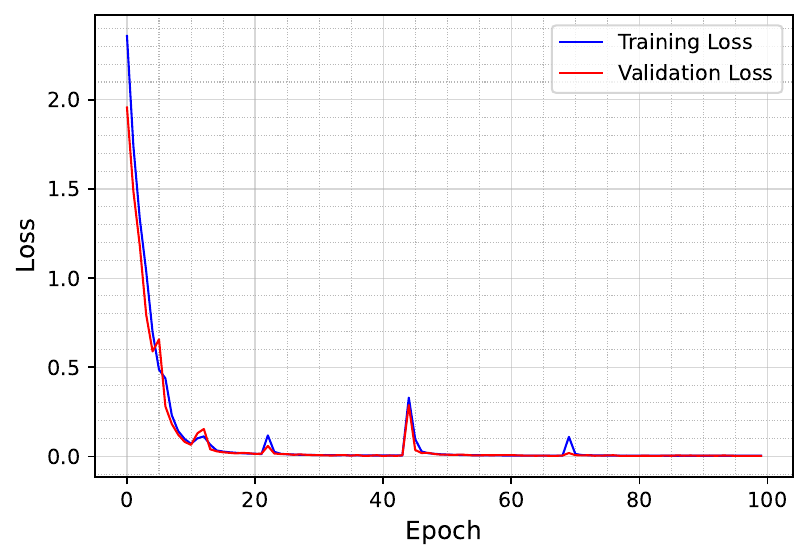}
        \caption{Training and validation loss over epochs}
        \label{fig:seq_gen_loss_eval}
    \end{subfigure}
    
    \caption{Performance evaluation of the LSTM-based RNN}
    \label{fig:eval_optimal_path_determiner}
    \vspace{-10pt}
\end{figure}

Before training the LSTM-based RNN to extract optimal attack sequences, we split the PEEVES dataset into training and validation sets with 80-20 ratio. In Fig. \ref{fig:eval_optimal_path_determiner}(a), we see the training and validation accuracy of the model over 100 epochs, while the training and validation loss are depicted in Fig. \ref{fig:eval_optimal_path_determiner}(b). We see that both training and validation accuracy becomes $>0.99$ after $epoch \negmedspace=\negmedspace 9$. Even though we see some unexpected decline in several occasions, for example, at $epoch \negmedspace = \negmedspace 44$, we still achieve $> \negmedspace 0.90$ accuracy all the time. On the other hand, both the training and validation loss becomes $\approx 0.02$ most of the time after $epoch \negmedspace = \negmedspace15$. Despite a few irregular loss at $epoch \negmedspace = \negmedspace 23, 44, 69$, we can conclude that after $epoch \negmedspace = \negmedspace 9$, the training and validation loss always stays $<\negmedspace 0.10$, and they become very close to $0$ a number of times. 


\vspace{-5pt}
\subsection{Rewards}
\begin{figure}[!h]
     \centering
     \captionsetup{justification=centering}
     \begin{subfigure}[t]{0.24\textwidth}
         \centering
         \includegraphics[width=\textwidth]{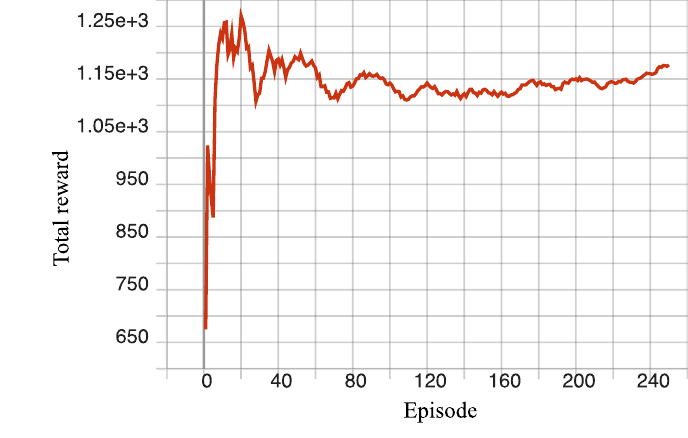}
         \caption{Reward over episodes}
         \label{fig:reward_in_each_episode}
     \end{subfigure}
     \hfill
     \begin{subfigure}[t]{0.24\textwidth}
         \centering
         \includegraphics[width=\textwidth]{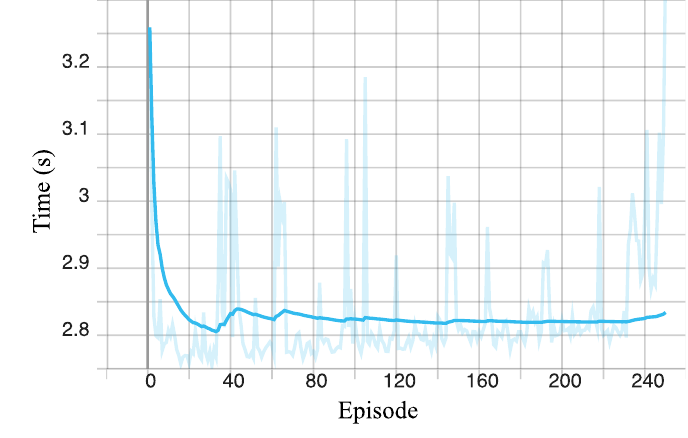}
         \caption{Time overhead over episodes}
         \label{fig:time_overhead}
     \end{subfigure}
     \hfill 
     \caption{Simulation results: reward and time overhead over episodes}
     \label{fig:reward_time_overhead}
\end{figure}

In our simulation, the defense agent is trained for $250$ episodes, interacting with the environment for $100$ epochs in each episode. Each interaction yields a discrete reward for the defense agent, and the defense agent accumulates all the rewards received over $100$ epochs together to compute the total reward it achieves in a single episode. Fig. \ref{fig:reward_time_overhead}(a) shows the total reward earned in our simulation over the entire $250$ episodes. We see that the total reward the defense agent receives for the first few episodes doesn't follow a stable pattern, and that's because instead of following a fixed optimal policy from the start, it explores all possible scenarios to determine the optimal policy that guarantees the maximization of the total reward at the end. Once the agent learns the optimal policy, the total reward it receives over the episodes follows a pretty stable pattern, as shown in Fig. \ref{fig:reward_time_overhead}(a) after $\approx 70$ episodes.

\subsection{Computation Overhead}
In our simulation, the defense agent dedicates most of its computation time to determine optimal state-action value pairs, sample experiences from the replay buffer, and train the policy network. Therefore, we compute the total time required in each episode to perform all these tasks and call it \textit{time overhead}. Fig. \ref{fig:reward_time_overhead}(b) shows how time overhead changes over the episodes. We can see that after $40$ episodes, the overhead reaches stability, close to $2.85$ seconds.



\subsection{Attack-Defense Dynamic}
\begin{figure}[!t]
    \centering
    \captionsetup{justification=centering}   
    \includegraphics[width=0.30\textwidth]{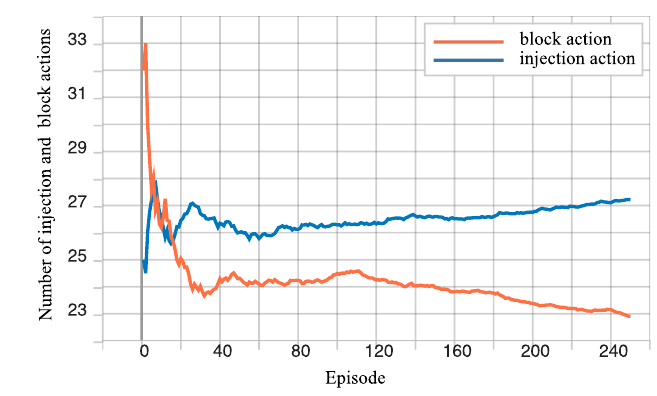}
    \caption{\small Number of injection and block actions over episodes}
    \label{fig:block-injection-dynamics}
    \vspace{-5pt}
\end{figure}

We design the defense agent to be reactive against the attack actions. If \textit{event injection ($a_1$)} actions are taken aggressively, the defense agent chooses to take \textit{blocking triggers ($a_4$)} actions more often. As we see in Fig. \ref{fig:block-injection-dynamics}, the defense agent initially becomes very aggressive, but it quickly starts to learn the attack-defense dynamic. Since \textit{blocking triggers} action negatively impacts the availability of the network devices, the defense agent avoids taking redundant \textit{blocking triggers} actions. Fig. \ref{fig:block-injection-dynamics} shows that the number of \textit{blocking triggers} actions taken by the defense agent is always smaller than the number of \textit{event injection} actions taken after a certain number of episodes ($\approx$ 30 episodes). It is possible that the attacker needs to inject multiple fake events to compromise a certain device, especially if any trigger operation to that particular device has been blocked earlier by the defense agent. Therefore, the objective of a trained defense agent is to take less \textit{blocking triggers} actions compared to the \textit{event injection} actions taken, which is clearly evident from Fig. \ref{fig:block-injection-dynamics}.

\vspace{-5pt}

\subsection{Impact of Injection Threshold}
\begin{figure}[!t]
    \centering
    \captionsetup{justification=centering}
    \includegraphics[width=0.29\textwidth] {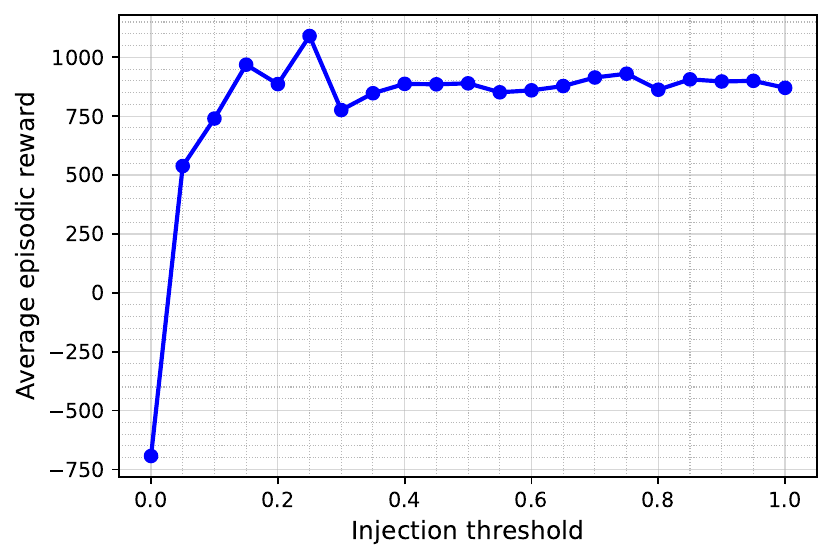}
    \caption{\small Average episodic return over injection thresholds}
    \label{fig:return_vs_injection_threshold}
    \vspace{-5pt}
\end{figure}

As we see in equation \eqref{eqn:reward_function}, the \textit{injection threshold}, parameterized as $k$, dictates the selection of optimal defense actions. To show the impact of $k$ on the reward function, we compute the \textit{average episodic reward}, $\Bar{R_t} = \frac{\sum_{t=1}^{N}R_t}{N}$, where $N = 100$ epochs, and $R_t$ represents the total reward achieved in a single epoch. In Fig. \ref{fig:return_vs_injection_threshold}, we show how the injection threshold $k$ ranging between $[0, 1]$ impacts the average episodic reward achieved by the defense agent. We see that after $k=0.25$, the impact of injection threshold on the reward function is quite constant, i.e., even though a greater $k$ allows an attacker to perform more \textit{event injection} operations, the defense agent still achieves a certain level of reward and effectively secure network nodes. In conclusion, once the defense agent determines the optimal defense policy, the increased aggressiveness in attack behavior barely changes the security status of the network.


\section{Conclusion}
In this paper, we propose a real-time defense system named \textit{IoTWarden} that infers attack behaviors upon an IoT network and counters a trigger-action attack by following an optimal action policy. We implement a neural network function approximator to select optimal action at each environment state by maximizing the security gain and train the defense policy using a Deep Q-Network. We implement the system using TensorFlow and conduct extensive simulation to evaluate the performance of the system. The experiment results show that the defense agent is capable of achieving stable security rewards with very low computation overhead by following an optimal defense policy under different aggressiveness levels of the injection operations conducted by an attacker.  As our approach focuses on real time detection, it can work in parallel with the static analysis-based security measures.

Future extensions of our approach contain the following aspects. First, we plan to improve our reinforcement learning-based mechanism so that it has user-configurable overhead. This will allow the mechanism to be adopted at different IoT environments with computation restriction. Second, we will extend our model with probability-based attack path and defense method selection so that end users have a better understanding of the coverage of the selected defense measures. Last but not least, we will deploy our approach at the SmartHome lab at our school and conduct experiments on real devices.   

\bibliographystyle{./bibliography/IEEEtran}
\bibliography{./bibliography/references}


\end{document}